\catcode`\@=11

\expandafter\ifx\csname draftcontrol\endcsname\relax\global\def\draftcontrol{0}
\fi

{\count255=\time\divide\count255 by 60
\xdef\hourmin{\number\count255}
\multiply\count255 by-60\advance\count255 by\time
\xdef\hourmin{\hourmin:\ifnum\count255<10 0\fi\the\count255}}
\def\draftdate{\number\month/\number\day/\number\year\ \ \ \hourmin }

\newcommand\makepapertitle{\par
  \begingroup
    \renewcommand\thefootnote{\@fnsymbol\c@footnote}%
    \def\@makefnmark{\rlap{\@textsuperscript{\normalfont\@thefnmark}}}%
    \long\def\@makefntext##1{\parindent 1em\noindent
            \hb@xt@1.8em{%
                \hss\@textsuperscript{\normalfont\@thefnmark}}##1}%
     \newpage
     \global\@topnum\z@   
     \@makepapertitle
     \thispagestyle{empty}\@thanks
  \endgroup
  \setcounter{footnote}{0}%
  \global\let\thanks\relax
  \global\let\makepapertitle\relax
  \global\let\@makepapertitle\relax
  \global\let\@thanks\@empty
  \global\let\@author\@empty
  \global\let\@date\@empty
  \global\let\@title\@empty
  \global\let\title\relax
  \global\let\author\relax
  \global\let\date\relax
  \global\let\and\relax
  \def\version{\let\version\@version\@gobble}
}
\def\@makepapertitle{%
  \newpage
   \ifnum\draftcontrol=1 {}
   \version\versionno
   \vskip 3em%
   \else
   \hfill\hbox to 3cm {\parbox{4cm}{\@pubnum}\hss}%
   \vskip 3em%
   \fi
   \begin{center}%
   \let \footnote \thanks
     {\LARGE \@title \par}%
     \vskip 1.5em%
     {\normalsize
       \lineskip .5em%
       \begin{center} 
         \@author
       \end{center} 
\par}%
     \vskip 1em%
     {\@bstract}%
     \end{center}%
     \vskip .5em
     \@date%
   \par
}

\gdef\@pubnum{}
\def\pubnum#1{%
  \gdef\@pubnum{#1}}

\gdef\@bstract{}
\def\Abstract#1{%
  \gdef\@bstract{%
   \parbox{\textwidth-0pc}{%
   \centerline{\bf Abstract}\penalty1000
   \noindent
   \renewcommand\baselinestretch{1.0}
   {#1}}}
}

\def\ps@paper{\let\@mkboth\@gobbletwo%
     \ifnum\draftcontrol=1
        \def\@oddfoot{\hbox to \textwidth{\tiny \versionno \hfil\tiny\draftdate}%
        \hskip -\textwidth \hbox to \textwidth{\hfil\rm\thepage\hfil}}%
     \else\def\@oddfoot{\hbox to \textwidth{\hfil\rm\thepage\hfil}}
     \fi
     \let\@evenfoot\@oddfoot
}

\def\body{\clearpage
          \pagestyle{paper}
        }
\newenvironment{acknowledgments}{%
\vskip 3.25ex
\noindent {\bf Acknowledgments}
}

\def\@version#1{\ifnum\draftcontrol=1
\typeout{}\typeout{#1}\typeout{}
\vskip3mm\centerline{\hbox{\fbox{\normalsize{\tt DRAFT -- #1 -- }
                   {\draftdate}}}}\vskip3mm
\fi}
\let\version\@version
\long\def\eqlabel#1{\ifnum\draftcontrol=1
                    \tag@false  
                    \tag*{(\theequation) \hbox to -0.2cm{\hspace{0cm}\small{#1}\hss}}
                    \refstepcounter{equation} 
                    \edef\@currentlabel{\theequation}
                    \ltx@label{#1}          
                    \else
                    \label{#1}
                    \fi
                    }
\let\st@bibitem\@bibitem
\let\st@lbibitem\@lbibitem
\ifnum\draftcontrol=1
  \def\@bibitem#1{%
    \st@bibitem{#1}\a@@label{#1}\ignorespaces}
  \def\@lbibitem[#1]#2{%
    \st@lbibitem[#1]{#2}\a@@label{#2}\ignorespaces}
  \def\a@@label#1{%
    \gdef\a@lab{\smash{\normalfont\small#1}}
    \ifvmode
      \if@inlabel
        \global\setbox\@labels\hbox{%
          \llap{\a@lab\let\a@lab\relax
                \kern\@totalleftmargin\kern\marginparsep}%
          \box\@labels}%
      \fi
    \fi}
\fi

\documentclass[12pt,letterpaper]{article}

\usepackage{amsmath,amssymb,array,calc,rotating,graphicx,psfrag}
\usepackage{hyperref}

\ifnum\draftcontrol=1
\fi

\renewcommand\baselinestretch{1.25}
\renewcommand\section{\@startsection {section}{1}{\z@}%
                                   {-3.5ex \@plus -1ex \@minus -.2ex}%
                                   {2.3ex \@plus.2ex}%
                                   {\normalfont\large\bfseries}}
\renewcommand\subsection{\@startsection{subsection}{2}{\z@}%
                                     {-3.25ex\@plus -1ex \@minus -.2ex}%
                                     {1.5ex \@plus .2ex}%
                                     {\normalfont\normalsize\bfseries}}
\renewcommand\subsubsection{\@startsection{subsubsection}{3}{\z@}%
                                     {-3.25ex\@plus -1ex \@minus -.2ex}%
                                     {1.5ex \@plus .2ex}%
                                     {\normalfont\normalsize\it}}

\numberwithin{equation}{section}

\def\projective   {{\mathbb P}}

\def\reals        {{\mathbb R}}

\def\revise#1       {\marginpar{\rule{2mm}{1cm} #1}}

\def\RR{\reals}
\def\PP{\projective}
\def\RP{\RR\PP}

\def\R{{\rm R}}

\def\sqr#1#2{{\vcenter{\vbox{\hrule height.#2pt  
 \hbox{\vrule width.#2pt height#1pt \kern#1pt
 \vrule width.#2pt}\hrule height.#2pt}}}}

\def\yboxit#1#2{\vbox{\hrule height #1 \hbox{\vrule width #1
\vbox{#2}\vrule width #1 }\hrule height #1 }}
\def\fillbox#1{\hbox to #1{\vbox to #1{\vfil}\hfil}}
\def\ybox{{\lower 1.3pt \yboxit{0.4pt}{\fillbox{8pt}}\hskip-0.2pt}}

\def\comments#1{}

\def\tr{{\rm tr}}

\def\CN{{\cal N}}

\def\P{\BP}

\def\II{\relax{I\kern-.10em I}}

\def\IZ{\relax\ifmmode\mathchoice
{\hbox{\cmss Z\kern-.4em Z}}{\hbox{\cmss Z\kern-.4em Z}}
{\lower.9pt\hbox{\cmsss Z\kern-.4em Z}}
{\lower1.2pt\hbox{\cmsss Z\kern-.4em Z}}\else{\cmss Z\kern-.4em
Z}\fi}
\def\IB{\relax{\rm I\kern-.18em B}}
\def\IC{{\relax\hbox{$\inbar\kern-.3em{\rm C}$}}}
\def\ID{\relax{\rm I\kern-.18em D}}
\def\IE{\relax{\rm I\kern-.18em E}}
\def\IF{\relax{\rm I\kern-.18em F}}
\def\IG{\relax\hbox{$\inbar\kern-.3em{\rm G}$}}
\def\IGa{\relax\hbox{${\rm I}\kern-.18em\Gamma$}}
\def\IH{\relax{\rm I\kern-.18em H}}
\def\II{\relax{\rm I\kern-.18em I}}
\def\IK{\relax{\rm I\kern-.18em K}}
\def\IP{\relax{\rm I\kern-.18em P}}

\def\Gslash{\relax G\kern-.6em \slash}

\def\inbar{\,\vrule height1.5ex width.4pt depth0pt}

\font\cmss=cmss10 \font\cmsss=cmss10 at 7pt
\def\IR{\relax{\rm I\kern-.18em R}}

\def\BR{\IR}
\def\BP{\IP}
\def\BR{\IR}
\def\BC{\IC}

\def\lp10{l_P^{10}}
\def\lp11{l_P^{11}}

\newcommand{\nc}{\newcommand}
\nc{\rnc}{\renewcommand}
\nc{\CY}{Calabi-Yau}
\nc{\CYM}{Calabi-Yau manifold}
\nc{\CYMs}{Calabi-Yau manifolds}
\nc{\DB}{D-Brane}
\nc{\DBs}{D-Branes}
\nc{\SUSY}{supersymmetry}
\nc{\Kah}{K\"ahler}
\nc{\cs}{complex structure}
\nc{\beq}{\begin{equation}}
\nc{\eeq}{\end{equation}}
\nc{\beqa}{\begin{eqnarray}}
\nc{\eeqa}{\end{eqnarray}}
\nc{\ntwo}{${\cal N}=2$}
\nc{\nOne}{${\cal N}=1$}
\nc{\hs}{\hspace{0.2in}}
\nc{\Z}{{\mathbb Z}}
\rnc{\P}{{\mathbb P}}
\rnc{\RP}{{\mathbb {RP}}}
\nc{\WP}{\mathbb{WP}}
\nc{\slag}{special Lagrangian}
\nc{\cn}{\C^n}
\nc{\rn}{\R^n}

\nc{\SO}{SO}
\nc{\Sp}{Sp}
\nc{\SU}{SU}

\nc{\Wtree}{W_{\mathrm tree}}
\nc{\Weff}{W_{\mathrm eff}}

\catcode`\@=12

\begin{document}

\title{\Large \bf Counting BPS Operators in the Chiral Ring of ${\cal N}=2$ Supersymmetric Gauge Theories\\
or\\
$\CN=2$ Braine Surgery}

\pubnum{%
hep-th/0611346\\
DIAS-STP-06-20}
\date{November 2006}

\author{
{Amihay Hanany$^\dag$, Christian R\"omelsberger$^{\dag,\#}$
}\\
\vspace{0.3 cm}
{\dag Perimeter Institute for Theoretical Physics\\
31 Caroline Street North\\
Waterloo Ontario N2L 2Y5\\
Canada
                   }\\
~\\
{\# Dublin Institute for Advanced Studies\\
10 Burlington Road\\
 Dublin 4\\
 Ireland}\\
~\\
ahanany@perimeterinstitute.ca, roemel@stp.dias.ie
}

\Abstract{This note is presenting the generating functions which count the BPS operators in the chiral ring of a $\CN=2$ quiver gauge theory that lives on N D3 branes probing an ALE singularity. The difficulty in this computation arises from the fact that this quiver gauge theory has a moduli space of vacua that splits into many branches -- the Higgs, the Coulomb and mixed branches. As a result there can be operators which explore those different branches and the counting gets complicated by having to deal with such operators while avoiding over or under counting. The solution to this problem turns out to be very elegant and is presented in this note. Some surprises with ``surgery" of generating functions arises.}

\enlargethispage{1.5cm}
\makepapertitle

\vfill \eject 
\tableofcontents

\body

\version\versionno

\section{Introduction}

\par
The chiral ring of a supersymmetric gauge theory is one of the fundamental properties one encounters in the study of 3+1 dimensional gauge theories. It reveals much of the structure of the vacuum of the theory as well as the spectrum of BPS operators. There are many examples in the literature in which the chiral ring is computed exactly, first classically and then with quantum corrections included. There are, however, many other cases in which the computation of the chiral ring is not possible since it involves many generators that satisfy relations which are not always easy to compute. At times one settles with specifying generic properties of the chiral ring like the dimension and the number of generators and at times the problem is too difficult to address.

A generic problem which deals with any supersymmetric gauge theory is the counting of operators in the chiral ring. This is a problem which is sometimes simpler than computing the chiral ring itself. For example, there are many situations in which the number of operators in the chiral ring is not changed with the inclusion of quantum effects and one can compute this number classically. Furthermore if we collect the number of operators into a generating function that counts the number of BPS operators which carry a collection of global $U(1)$ charges then this generating function encodes some simple properties of the supersymmetric gauge theory, properties like the dimension of the moduli space of vacua, the number of generators in the chiral ring and the number of relations they satisfy. The generating function also encodes the ``volume" of the moduli space of vacua.

Counting problems of operators in the chiral ring of a supersymmetric gauge theory has recently attracted some attention due to few reasons.
\begin{enumerate}

\item Better understanding of supersymmetric gauge theories \cite{Pouliot:1998yv} and of a class of string theory backgrounds \cite{Benvenuti:2006qr, Noma:2006pe,Oota:2006eg, Lee:2006hw}.

\item The AdS/CFT correspondence -- a check of this by doing the computation on the gauge theory side and comparing with the computation from the gravity side using giant gravitons \cite{Biswas:2006tj, Sinha:2006sh} or dual giant gravitons \cite{Mandal:2006tk, Martelli:2006vh, Basu:2006id}.

\item Some related work on the BPS index in quiver gauge theories \cite{Romelsberger:2005eg, Kinney:2005ej, Nakayama:2005mf, Nakayama:2006ur}.

\item Possible microstate counting of supersymmetric black holes.

\end{enumerate}
In this paper we improve the generic counting problem, which is still a very hard problem, by introducing the method of surgery and by solving for a large class of supersymmetric gauge theories.

\section{Statement of the Problem}

The paper \cite{Benvenuti:2006qr} solves the problem of counting BPS operators for a large class of quiver gauge theories. This is done essentially by exploiting the known geometric properties of the moduli space of vacua for these theories. The class of theories which are addressed in this work all have their moduli space of vacua as Calabi Yau (CY) spaces or their symmetric products. There are, however, many cases in which this work does not cover the complete counting and avoids questions of mixed branches and the emergence of fractional branes in the geometry. The present work aims at addressing these issues. We are going to study a class of ${\cal N}=2$ supersymmetric gauge theories in 3+1 dimensions which arise on the world volume of a D3 brane probing an ALE singularity of type A, D, or E. These theories have a moduli space of vacua which consists of a Higgs branch, a Coulomb branch, and more importantly mixed branches. Counting the BPS operators on the Higgs branch of this theory is relatively easy and was done in \cite{Benvenuti:2006qr}. Counting the BPS operators on the Coulomb brach follows the same techniques as in \cite{Benvenuti:2006qr} and is a simple application there. It will be reviewed in section \ref{coulomb}. Counting BPS operators in the mixed branch has not been done and is the focus of the present paper. The plan of the paper is as follows. In section \ref{preliminary} we introduce some notation. In section \ref{functions} we introduce the various generating functions on the Higgs, Coulomb and mixed branches. We discuss the intersections between these branches and end by proposing the right solution of the problem. Equation \eqref{Mnu} is the main result of the paper. In section \ref{surgery} we introduce the concept of surgery which is the main tool that allows the derivation of the main result. In section \ref{examples} we present some examples and then conclude. 

Note that we are doing the calculations in the weakly coupled theory, where we can distinguish between chiral operators which just come from matter fields and chiral operators which contain gauginos. In this paper we will not deal with the gauginos.

\subsection{Some preliminaries and notation}
\label{preliminary}

We will be interested in $N$ D3 branes living in Type IIB on $\RR^{3,1}\times \RR^2\times$ ALE, where the D3 branes fill $\RR^{3,1}$ and the ALE space can admit any of the A, D, or E singularities, namely it looks like $\BC^2/\Gamma$ and $\Gamma$ can be any of the cyclic, dihedral, or the exceptional discrete subgroups of $SU(2)$.
In \cite{Benvenuti:2006qr} the generating function counting single trace BPS operators in the chiral ring is denoted by $f$ and the generating function counting multi trace BPS operators in the chiral ring is denoted by $g$. In this paper we do not deal with the distinction between single trace and multi trace operators and just concentrate on multi trace operators, with the understanding that single trace operators are always included among the multi trace operators. Since these functions depend on the number of branes, we will put a subscript $g_N$ for counting BPS operators for $N$ D3 branes.

The global symmetry for the A series contains a $U(1)^3$ subgroup (including R charges) since these spaces are toric manifolds. The global symmetry for the D and E series contains a $U(1)^2$ subgroup. Correspondingly we will denote the chemical potentials for these charges by $t_1, t_2, t_3$, which are taken to be arbitrary complex numbers living in the unit disc. For the D and E cases $t_1$ and $t_2$ will be replaced by one parameter which will be denoted by $t$. The parameter $t_3$ counts operators which carry charge associated with the complex line in the background called $\RR^2$ above.

The generating functions do depend on the background and therefore will carry this information in the following manner
\beq
g_N(t_1, t_2, t_3 ; \BC^2/\Gamma\times\BC).
\eeq
To simplify this notation we will abbreviate by writing
\beq
g_N(t_1,t_2,t_3; \Gamma),
\eeq
for cases when there is no ambiguity. The functions $g$ admit a Taylor expansion around the origin in the parameters $\{t_i\}$ and their coefficients are positive integers which count the number of BPS operators. For example the expansion
\beq
g(t_1,t_2, t_3) = \sum_{i=0}^\infty\sum_{j=0}^\infty\sum_{k=0}^\infty d_{ijk} t_1^i t_2^j t_3^k
\eeq
indicates that $d_{ijk}$ is the number of multi trace BPS operators in the chiral ring, which carry charges $i$, $j$, and $k$ with respect to the three global $U(1)$ charges, etc.

The moduli space of vacua of the gauge theory contains a collection of branches, the Higgs branch, the Coulomb branch and mixed branches. On each of these branches there will be different BPS operators, some which exist in few branches and some which exist only on a specific branch. Correspondingly, there will be generating functions which count BPS operators in each branch. We will replace the notation $g$ by the first letters of these branches, hence we will have $h_N$, $c_N$ which are generating functions counting multi trace BPS operators in the chiral ring of the Higgs and Coulomb branches, respectively.

We can further introduce a chemical potential $\nu$ for the number of branes, $N$.
Using this parameter the generating functions for counting multi trace BPS operators on the Higgs branch is collected into a single generating function called $H$ (for Higgs) which has the following expansion
\beq
H ( \nu ; t_1,t_2,t_3; \Gamma) = \sum_{N=0}^\infty \nu^N h_N (t_1,t_2,t_3; \Gamma).
\eeq
We can similarly collect the generating functions for multi trace operators in the Coulomb branch under
\beq
C ( \nu ; t_1,t_2,t_3; \Gamma) = \sum_{N=0}^\infty \nu^N c_N (t_1,t_2,t_3; \Gamma).
\eeq

\section{The elementary partition functions}
\label{functions}

In this section we derive the building blocks for the full solution of the problem: The partition function for the Higgs branch, the Coulomb branch and the intersection between those.  In the next section we will then use surgery to combine those building blocks into a full solution.

\subsection{The Higgs Branch}\label{higgs}

The Higgs branch of the gauge theory on a single D3 brane probing an ALE singularity is a copy of the complex line $\BC$ times the ALE space. When $N$ D3 branes probe this singularity the moduli space is a copy of the $N$-th symmetric product of $\BC$ times the ALE. This property makes it particularly simple to write down the generating functions on the Higgs branch by simply exploiting the properties of the plethystic exponential as discussed in detail in \cite{Benvenuti:2006qr}.
The resulting generating function $H$ can be written in terms of a single function, $h_1$, counting multi trace BPS operators for a single D3 brane as
\beq
H(\nu ; t_1, t_2, t_3 ; \Gamma ) =  \exp\biggl(\sum_{k=1}^\infty\frac{\nu^kh_1(t_1^k,t_2^k,t_3^k ; \Gamma)}{k}\biggr).
\label{Hnu}
\eeq
The computation of $h_1$ is further simplified by noting that the ALE spaces are manifolds of complete intersections and their function follows immediately from the possible symmetries of the defining equations. These functions coincide with the so called ``Molien Invariant" of the discrete group $\Gamma$ and we recall from \cite{Benvenuti:2006qr}:
\beqa
h_1(t_1,t_2,t_3; \Z_n) &=& \frac{1-t_1^nt_2^n}{(1-t_1^n)(1-t_2^n)(1-t_1t_2)(1-t_3)},\\
h_1(t_1,t_2,t_3; D_{n+2}) &=& \frac{1+t^{2n+2}}{(1-t^4)(1-t^{2n})(1-t_3)},\\
h_1(t_1,t_2,t_3; E_6) &=& \frac{1-t^{24}}{(1-t^6)(1-t^8)(1-t^{12})(1-t_3)},\\
h_1(t_1,t_2,t_3; E_6) &=& \frac{1-t^{36}}{(1-t^8)(1-t^{12})(1-t^{18})(1-t_3)},\\
h_1(t_1,t_2,t_3; E_6) &=& \frac{1-t^{60}}{(1-t^{12})(1-t^{20})(1-t^{30})(1-t_3)},
\eeqa

The trivial dependence on $t_3$ indicates the fact that it counts operators on the complex line.

\subsection{The Coulomb Branch}
\label{coulomb}

${\cal N}=2$ supersymmetric gauge theories have a single adjoint valued complex field and are well studied theories. For the purpose of counting BPS operators in the chiral ring we can think of symmetric functions of the eigenvalues of the adjoint matrix. Alternatively, we can think of all independent Casimir invariants, the number of which is equal to the rank of the gauge group. For a gauge group of rank $r$ the moduli space of vacua is freely generated by the first $r$ Casimir invariants or, alternatively, by the lowest $r$ symmetric functions of the eigenvalues (a special discussion needs to be made for exceptional groups but this is not going to be of concern to us in the following.) Restricting to a $U(r)$ gauge group, the moduli space of vacua ${\cal M}_r$ is a copy of the symmetric product of the complex line,
\beq
{\cal M}_r = S_r(\BC) = \frac{\BC^r}{S_r},
\label{Mr}
\eeq
where $S_r$ is the symmetric group in $r$ elements.
The generating function for the BPS chiral operators of this moduli space was easily calculated in \cite{Benvenuti:2006qr} to be
\beq
c_r(t_3 ; \BC ) = c_1 (t_3 ; S_r(\BC) ) = \prod_{i=1}^r \frac{1}{(1-t_3^i)}.
\label{gr}
\eeq
Next, suppose we are dealing with a product of gauge groups,  $\prod_j U(r_j)$ as is the typical case for a quiver gauge theory. Then the moduli space of vacua in the Coulomb branch is easily generalized to be
\beq
{\cal M}_{\{r_j\}} = \prod_j S_{r_j} (\BC) = \prod_j \frac{\BC^{r_j}}{S_{r_j}},
\eeq
and by analogy the generating function counting BPS operators in the chiral ring is computed to be
\beq
c_{\{r_j\}}(t_3 ; \BC ) = \prod_j c_1 (t_3 ; S_{r_j}(\BC) ) = \prod_j \prod_{i_j=1}^{r_j} \frac{1}{(1-t_3^{i_j})}.
\label{gencoul}
\eeq
We are now ready to apply this discussion to the case of the ALE singularities.
The gauge group $G(\Gamma)$ living on $N$ D3 branes probing an ALE singularity is
\beqa
G (\Z_n) & = &U(N) ^ n \\
G (D_{n+2}) & = & U(N)^4 U(2N) ^{n-1} \\
G (E_6) & = & U(N)^3 U(2N)^3 U(3N) \\
G (E_7) & = & U(N)^2 U(2N)^3 U(3N)^2 U(4N) \\
G (E_8) & = & U(N) U(2N)^2 U(3N)^2 U(4N)^2 U(5N) U(6N)
\eeqa
It should be pointed out that the different gauge group factors represent fractional branes which are many more than the physical $N$ D3 branes. As a result each type of fractional brane carries a conserved charge which is associated to its fractional number. For example, the $n$ different fractional charges in the $\Z_n$ singularity are such that taking the sum of precisely one of each of type of these fractional branes produces a physical brane. In the equation for the generating function \eqref{gencoul} we are taking the chemical potential $t_3$ to be the same for all types of fractional branes. It is possible to keep track of each of these fractional brane charges by taking a larger set of chemical potentials, say $q_j$, each of which carries information about the specific fractional brane charge. Such chemical potentials will need to obey a conservation law which implies that for the example of $\Z_n$, the sum of $n$ fractional branes is a physical brane but for simplicity we will assume that all fractional branes carry the same charge, $t_3$.

Using this gauge theory data written above we can compute the generating function $c_N(t_3 ; \Gamma) $ for counting BPS operators in the Coulomb branch of $N$ D3 branes probing an ALE singularity of type $\Gamma$ to be
\beqa\label{gencoulZ}
c_N (t_3 ; \Z_n) & = & \prod_{i=1}^N \frac{1}{(1-t_3^i)^n} \\
c_N (t_3 ; D_{n+2}) & = & \prod_{i=1}^N \frac{1}{(1-t_3^i)^4} \prod_{i=1}^{2N} \frac{1}{(1-t_3^i)^{n-1}} \\
c_N (t_3 ; E_6) & = & \prod_{i=1}^N \frac{1}{(1-t_3^i)^3}\prod_{i=1}^{2N} \frac{1}{(1-t_3^i)^3}\prod_{i=1}^{3N} \frac{1}{(1-t_3^i)} \\
c_N (t_3 ; E_7) & = & \prod_{i=1}^N \frac{1}{(1-t_3^i)^2}\prod_{i=1}^{2N} \frac{1}{(1-t_3^i)^3}\prod_{i=1}^{3N} \frac{1}{(1-t_3^i)^2}\prod_{i=1}^{4N} \frac{1}{(1-t_3^i)} \\
c_N (t_3 ; E_8) & = & \prod_{i=1}^N \frac{1}{(1-t_3^i)}\prod_{i=1}^{2N} \frac{1}{(1-t_3^i)^2}\prod_{i=1}^{3N} \frac{1}{(1-t_3^i)^2}\prod_{i=1}^{4N} \frac{1}{(1-t_3^i)^2}\prod_{i=1}^{5N} \frac{1}{(1-t_3^i)}\prod_{i=1}^{6N} \frac{1}{(1-t_3^i)} \nonumber
\label {gencoulE8}
\eeqa
We can further collect the generating functions depending on the number of branes $N$ into a single generating function $C$ (for Coulomb) but it is not known to us how to sum this function into a simpler form. We will settle with an implicit expression,
\beq
C (\nu ; t_3 ; \Gamma ) = \sum_{N=0}^\infty \nu^N c_N (t_3 ; \Gamma ),
\label{Cnu}
\eeq
with $c_N$ expressed in equations (\ref{gencoulZ}) - (\ref{gencoulE8}).
The case $n=1$ in equation \eqref{gencoulZ} corresponds to flat space and in fact to the ${\cal N}=4$ supersymmetric gauge theory restricted to one adjoint field. It takes the form (since the group $\Gamma$ is trivial we replace the notation to be $\BC^2$ corresponding to flat space)
\beq
C (\nu ; t_3 ; \BC^2 ) = \sum_{N=0}^\infty \nu^N \prod_{i=1}^N \frac{1}{(1-t_3^i)} = \prod_{k=0}^\infty \frac{1}{(1-\nu t^k)},
\eeq
and is related to the number of integer partitions. (For $\nu=1$ this is the inverse of the Euler function.)
The case $n=1$ is in fact relevant to the discussion on the mixed branch to which we now turn.

\subsection{The Intersection between Coulomb and Higgs Branches}

It is useful to think about the intersection between the Coulomb and Higgs branches by using some physical reasoning. Let us turn to branes as they serve as a good intuitive thinking about such points. The Coulomb branch meets the Higgs branch at the origin of the ALE space in the Higgs branch. 
On the other hand, the Coulomb branch is the moduli space of fractional branes and in order to move to the Higgs branch they must split into appropriate sets of coinciding fractional branes that form a D3-brane bound state. At the intersection of both branches the physical brane is still free to move along the complex direction corresponding to $\RR^2$ and with chemical potential $t_3$. When $M$ physical branes reach the intersection between the Higgs and Coulomb branches they are indistinguishable and therefore their moduli space of vacua is that of equation (\ref{Mr}),
\beq
{\cal M}_M = S_M(\BC) = \frac{\BC^M}{S_M},
\label{MM}
\eeq
with a generating function which is familiar by now, as in equation (\ref{gr})
\beq
l_M(t_3 ; \BC ) = l_1 (t_3 ; S_M(\BC) ) = \prod_{i=1}^M \frac{1}{(1-t_3^i)}.
\label{gM}
\eeq
We can collect all of these functions into a single generating function of the complex line (hence the name $L$) using the chemical potential, $t_3$, for the number of such branes, $M$, at the intersection between the Coulomb and Higgs branches. (It is important to note that the number of branes at the intersection, $M$, is different than the total number of D3 branes, $N$, since not all branes need to be at the intersection.)
\beq
L (\nu ; t_3 ) = \sum_{M=0}^\infty \nu^M \prod_{i=1}^M \frac{1}{(1-t_3^i)} = \prod_{k=0}^\infty \frac{1}{(1-\nu t_3^k)} = \exp\biggl(\sum_{k=1}^\infty\frac{\nu^k}{k(1-t_3^k)}\biggr),
\label{Lnu}
\eeq
We now turn to final words about the mixed branch.

\subsection{The Mixed Branches}
\label{mixed}

Generically, there are many mixed branches in which there are $M$ D3-branes in the Higgs branch and the rest of the branes are in the Coulomb branch. Away from special points the moduli space of such a mixed branch looks like the direct product of the appropriate Higgs and Coulomb branch.  

The generating function for BPS operators of  branes propagating on the Higgs branch is given by $H$, equation (\ref{Hnu}), and the generating function for BPS operators of branes propagating along the Coulomb branch is given by $C$, equation (\ref{Cnu}). 
Finally, we should be careful not to over-count BPS operators of branes which live at the intersection between the Coulomb and Higgs branches, as they are counted in both. These come with a generating function given by $L$. In the next section we will show that, using the rules of ``surgery", we can simply combine these points and are now ready to state the main result of this paper: the generating function for counting BPS operators in the mixed branch is given by the product of the Higgs and Coulomb branch generating functions divided by the generating function of their intersection.
\beq
G(\nu; t_1, t_2, t_3; \Gamma) =\frac { H(\nu; t_1, t_2, t_3; \Gamma) C(\nu;  t_3; \Gamma) } { L(\nu; t_3)}.
\label{Mnu}
\eeq
Even though this result is very simple to state and looks intuitive,
it is harder to prove. This is the main point of the next section.

\section{Surgery}
\label{surgery}

The goal of this section is to prove equation \eqref{Mnu} with the help of surgery techniques which we develop using examples in sections \ref{example1} and \ref{example2}. 

\subsection{An illustrative Example}\label{example1}

The simplest example for a situation in which there is a moduli space of solutions to an algebraic equation which has different branches is probably the equation for two complex variables $x$ and $y$,
\beq\label{xyzero}
xy=0.
\eeq
The solution to this equation has two branches:
\begin{enumerate}
\item $x\not =0$, $y=0$ -- The moduli space is a copy of the complex line.
\item $x=0$, $y\not =0$ -- Here by symmetry the moduli space is again a copy of the complex line.
\item The two branches meet at a single point, $x=y=0$ -- it can be termed the ``mixed branch" even though it is degenerate but this is good enough for the purpose of our demonstrative example.
\end{enumerate}
The chiral operators can be identified with all the polynomials in $x$ and $y$ modulo the relation (\ref{xyzero}). In this example it is not hard to count these polynomials explicitly: The relation implies that every monomial either contains only $x$ or only $y$. Let us denote by $t_1 (t_2)$ the parameter which counts the number of $x$'s ($y$'s).
The monomials of the form $x^n$ ($n\ge 1$) are counted by $\frac{t_1}{1-t_1}$. Similarly, the monomials of the form $y^n$ ($n\ge 1$) are counted by $\frac{t_2}{1-t_2}$. Furthermore, there is the identity operator. In summary, the generating function to count the chiral operators is
\beq
g(t_1,t_2)=\frac{t_1}{1-t_1}+\frac{t_2}{1-t_2}+1=\frac{1-t_1t_2}{(1-t_1)(1-t_2)}.
\label{straight}
\eeq
This counting problem can be translated into a language of moduli spaces:  The chiral operators correspond to holomorphic functions on the moduli space. For this reason we need to count all the holomorphic functions on the moduli space.

There are infinitely many holomorphic functions on the complex line, $x^n$, $n=0, 1, 2, \ldots$. The generating function for these functions on this moduli space can be described as follows.
For the first branch, $x$ is a good coordinate and the generating function of the line is 
$\frac{1}{1-t_1}$. By a complete analogy the generating function on the second branch is $\frac{1}{1-t_2}$. The two branches meet at the origin and therefore we should be careful not to double count this point. Alternatively the identity function appears in both generating functions and we should not double count it. As a result the total generating function is
\beq
g(t_1,t_2) = \frac{1}{1-t_1}+\frac{1}{1-t_2}-1 = \frac{1-t_1 t_2}{(1-t_1)(1-t_2)},
\eeq
reaching the same result as in equation \eqref{straight}. This is the simplest type of ``surgery" of moduli spaces. We add the contributions from the two branches and subtract the contribution from the ``mixed" branch.

Let us derive this generating function using an third derivation which follows from the reasoning given in \cite{Benvenuti:2006qr}. Inspection of the equation $xy=0$ reveals that this is a simple example of a complete intersection manifold, namely one can treat this as a one complex dimensional space given as one relation in two variables. As such it satisfies the condition that the dimension of the moduli space of solutions and the number of relations sum up to the number of variables, 1+1 = 2. As such the moduli space qualifies as a complete intersection. We recall that for such a case the computation of the generating function requires to identify the $U(1)$ isometries and their weights. We can observe that there are two $U(1)$ isometries to this equation: counting $x$'s this gives a weight 1 to $x$, 0 to $y$ and the relation has a weight 1. The second $U(1)$ counts the number of $y$'s giving weight 0 to $x$, 1 to $y$ and the relation has weight 1. These charges can be summarized in Table \ref{default}.

\begin{table}[htdp]
\caption{$U(1)$ charges for the equation $xy=0$}
\begin{center}
\begin{tabular}{|c||c|c||c|}
\hline
Variable & x & y & relation \\ \hline
$t_1$ & 1 & 0 & 1\\ \hline
$t_2$ & 0 & 1 & 1\\ \hline
\end{tabular}
\end{center}
\label{default}
\end{table}

Collecting these together we can write the generating function as
\beq
g(t_1,t_2) = \frac{1-t_1 t_2}{(1-t_1)(1-t_2)},
\eeq
where the numerator encodes the weight of the relation and the denominator denotes the weight of the generators $x$ and $y$.
Having seen the three methods of computing the same generating function it turns out that the second method, which involves the simplest type of ``surgery", will be our main tool. Let us see this in the following example.

\subsection{A more realistic example}\label{example2}

A more 
interesting model is a Wess-Zumino model with three chiral multiplets $x$, $y$ and $z$ and a superpotential
\beq
W=xyz.
\label{Wxyz}
\eeq
This model arises as a degenerate limit of a gauge theory. It is the theory on a single D3 brane which probes a $\BC^3/(\Z_n\times\Z_n)$ singularity with a non-trivial discrete torsion turned on \cite{Douglas:1999hq}.
For such a theory the superpotential comes with a coefficient $(1-\epsilon)$, with $\epsilon$ a root of 1 but this coefficient does not affect our discussion.
The superpotential \eqref{Wxyz} imposes three relations,
\beq
yz=xz=xy=0.
\label{relxyz}
\eeq
here the moduli space has three branches:
\begin{itemize}
\item $x\not=0$, $z=y=0$,
\item $y\not=0$, $x=z=0$,
\item $z\not=0$, $x=y=0$,
\item The three branches meet at a single point $x=y=z=0$, the ``mixed branch".
\end{itemize}
A similar reasoning as in the previous section leads  to
\beqa
g(t_1,t_2,t_3)&=&\frac{t_1}{1-t_1}+\frac{t_2}{1-t_2}+\frac{t_3}{1-t_3}+1\\&=&\frac{1}{1-t_1}+\frac{1}{1-t_2}+\frac{1}{1-t_3}-2\\&=&\frac{1-t_1t_2-t_1t_3-t_2t_3+2t_1t_2t_3}{(1-t_1)(1-t_2)(1-t_3)}.
\label{gxyz}
\eeqa
Here the second equality follows from ``surgery" and since we counted the ``mixed branch" three times we need to subtract 2 in order to avoid over counting.
This moduli space turns, however, not to be a complete intersection. This can be seen by observing that
\beq
(xy)z=x(yz)=y(xz),
\label{syzxyz}
\eeq
i.e. there are relations of relations.
In fact taking the Plethystic Logarithm of the generating function $g$ in equation \eqref{gxyz} gives an infinite series, showing that indeed the moduli space is not a complete intersection and has an expansion
\beq
PL[g(t_1,t_2,t_3)] :=\sum_{k=1}^\infty \frac{\mu(k) g(t_1^k,t_2^k,t_3^k)}{k} = (t_1+ t_2 + t_3) - (t_1t_2 + t_2t_3 + t_3t_1)+2t_1t_2t_3+\ldots
\eeq
Here $\mu$ is the M\"obius function.
The first contribution indicates that there are 3 generators for the chiral ring, $x, y, z$. It appears with a positive sign. The second contribution indicates that there are 3 relations, as in equation \eqref{relxyz}. It appears with a negative sign. The third contribution indicates the relations between relations and indeed there are two of them as in equation \eqref{syzxyz}. It appears with a positive sign consistent with the fact that for each relation there is an extra minus sign. The remaining terms in the expansion indicate an increasingly complicated pattern of relations for relations (syzygies).

\subsection{The quiver gauge theory}

As described in section \ref{mixed}, the moduli space of a $\CN=2$ supersymmetric quiver gauge theory splits into many different branches, which are mixtures of the Higgs and the Coulomb branch. Inspired by the D-brane picture, we develop here a diagrammatic way to describe the moduli space. For simplicity we restrict ourselves to an $A_1$ singularity and often to a representative diagram instead of writing out the most general expression. It should however be clear how to generalize this. The final result equation (\ref{finresult}) is independent of those details.

We begin with some notation.
A diagram of the form
\beq
\raisebox{-.15in}{\includegraphics[width=1.2in]{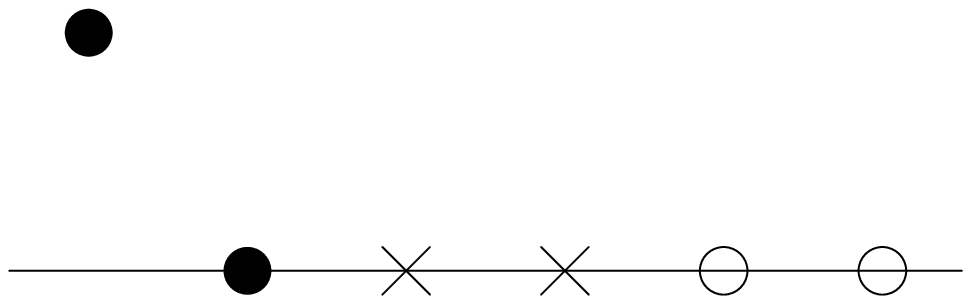}}
\eeq
can be interpreted as follows:
\begin{enumerate}
\item The horizontal axis is the $\BR^2$ direction and is counted by the parameter $t_3$.
\item The vertical axis is the direction in the ALE space. 
\item A filled dot off the horizontal axis denotes a D3-brane that can move around the ALE space.
\item A filled dot sitting on the horizontal axis denotes a D3-brane that is stuck at the origin of the ALE space and is indistinguishable from another D3-brane that happens to sit at the origin of the ALE space.
\item On an $A_1$ singularity there are two types of fractional branes which are stuck to the singular point. A cross denotes a fractional D3-brane of the first kind.
\item An empty circle denotes a fractional D3-brane of the second kind. When a brane of the first kind and a brane of the second kind combine they form a physical D3 brane which can move away to the ALE space.
\end{enumerate}
From this picture it is clear that if a cross and an empty circle are at the same position, they can be replaced by a filled dot at the same position. This property poses the main difficulty in counting the chiral operators.

\subsubsection{The alternating sum}

First we note that the full moduli space can be split up into Higgs, Coulomb and mixed branches.
The $m$-th mixed branch, $B_m$, has $m$ D3-branes exploring the ALE space and the other $N-m$ D3-branes split up into fractional branes. An example is
\beq
B_2: \raisebox{-.15in}{\includegraphics[width=1.2in]{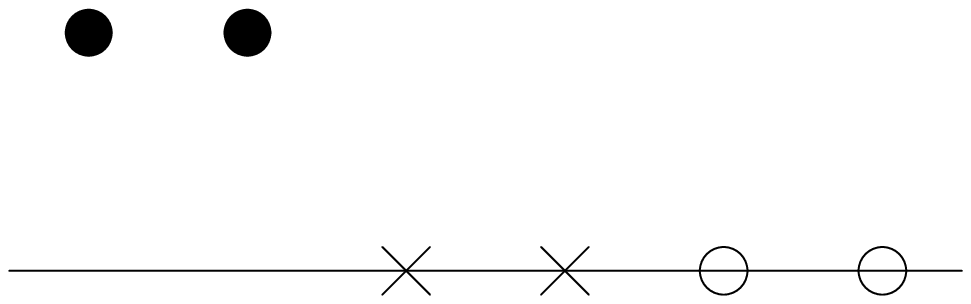}}
\eeq

Also, there is a natural notion of neighboring branches. The neighboring branches $B_m$ and $B_{m+1}$ intersect in a locus $I_m$, where $m$ D3-branes are exploring the ALE space, one more D3-brane is stuck at the origin of the ALE space and the remaining $N-m-1$ D3-branes split into fractional D3-branes. Here is a diagrammatic representation of $I_1$.
\beq
I_1: \raisebox{-.15in}{\includegraphics[width=1.2in]{graph27.eps}}
\eeq

When we glue two neighboring branches together, we have to add their generating functions for chiral primaries, but then we have to subtract the generating function for chiral primaries in the intersection locus. Stepwise we can continue gluing on more neighboring branches. It is not hard to see that if we glue another neighboring branch, $B_{m+l+1}$, to a bunch of neighboring branches, $B_m, \cdots, B_{m+l}$, the intersection is actually $I_{m+l}$. This leads to the alternating sum which is written here for the case of $N=3$,
%
\beq\label{altsum}
g=\raisebox{-.15in}{\includegraphics[width=.51in]{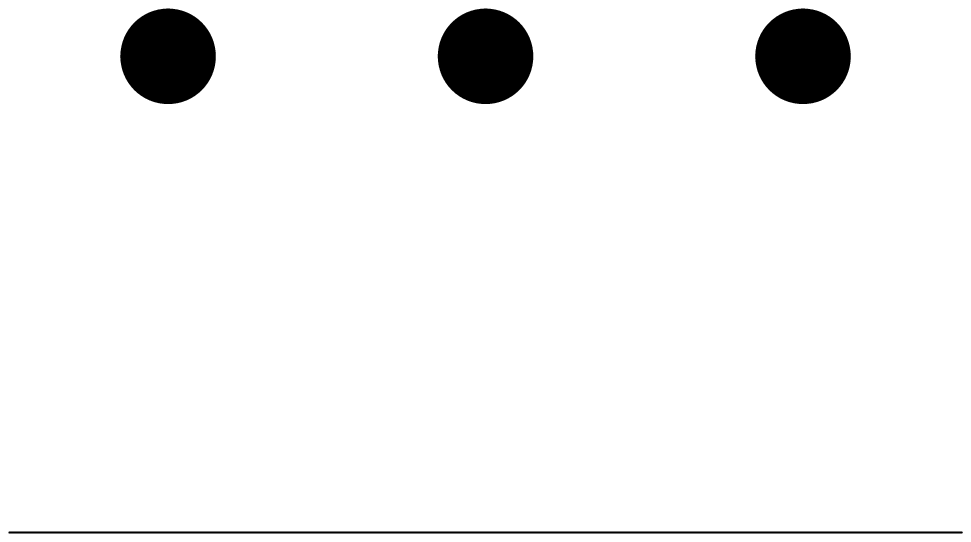}}
-\raisebox{-.15in}{\includegraphics[width=.51in]{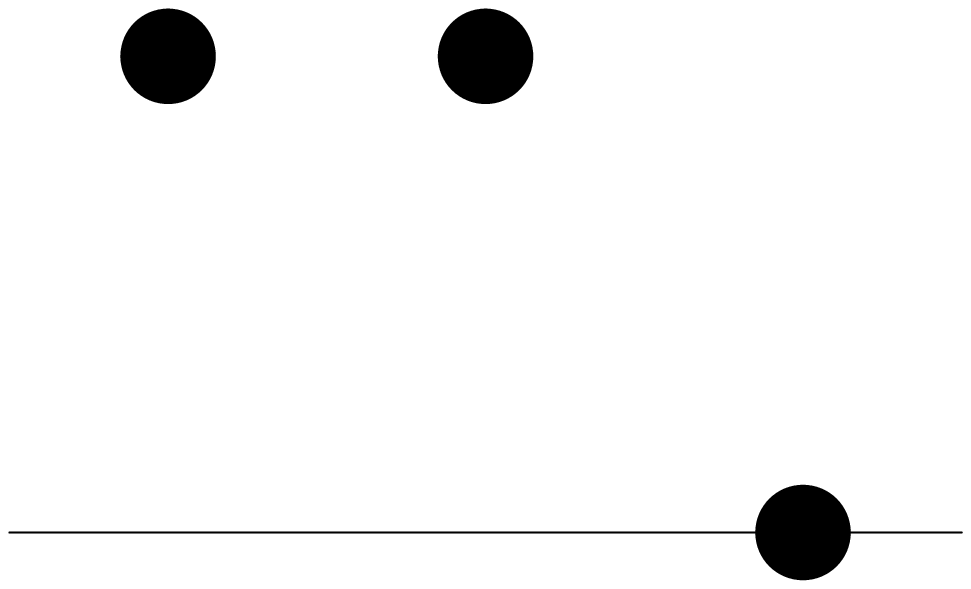}}
+\raisebox{-.15in}{\includegraphics[width=.68in]{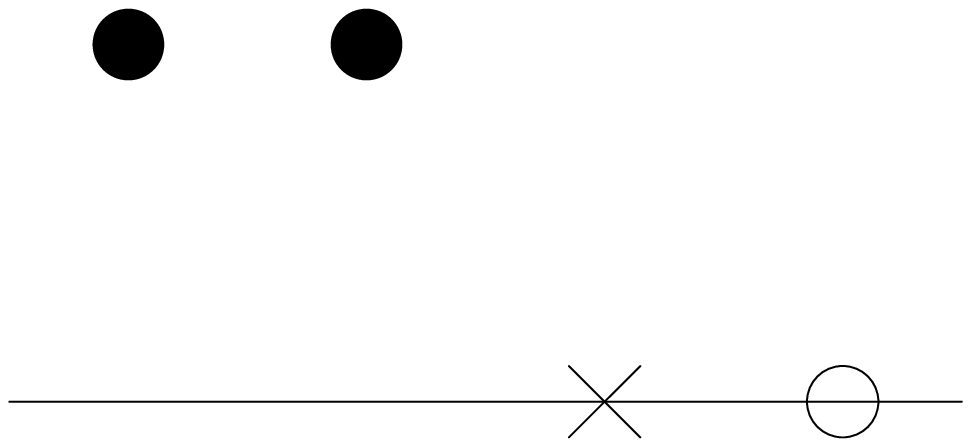}}
-\raisebox{-.15in}{\includegraphics[width=.68in]{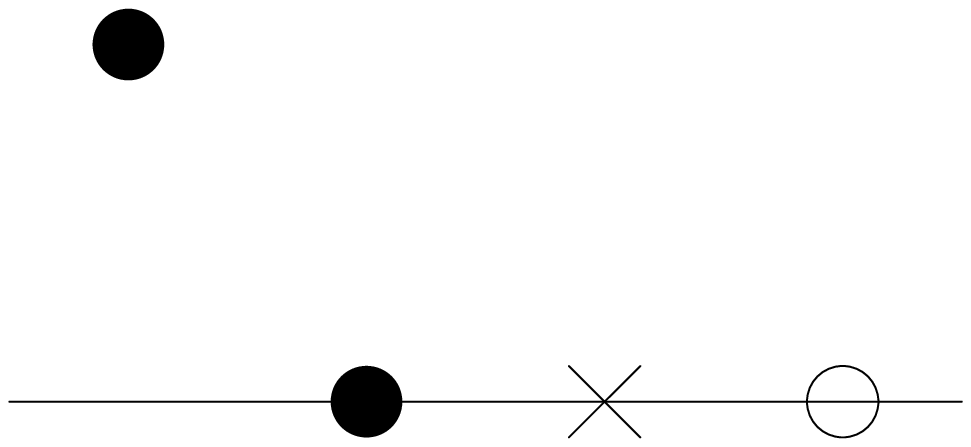}}
+\raisebox{-.15in}{\includegraphics[width=.85in]{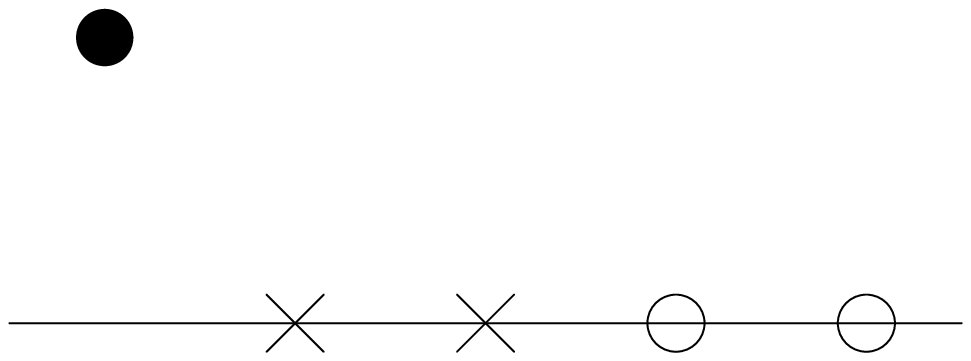}}
-\raisebox{-.15in}{\includegraphics[width=.85in]{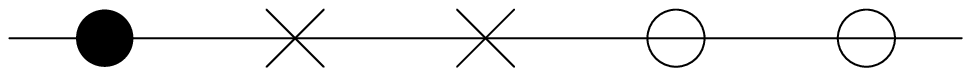}}
+\raisebox{-.15in}{\includegraphics[width=1.02in]{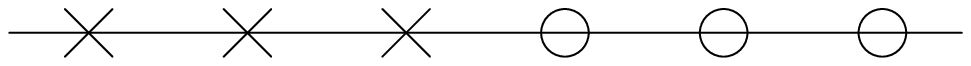}}
\eeq

Most terms like
\beq
\raisebox{-.15in}{\includegraphics[width=1.2in]{graph21.eps}} \qquad{\rm or}\qquad
\raisebox{-.15in}{\includegraphics[width=1.2in]{graph27.eps}}
\eeq
in this alternating sum are still quite complicated because fractional D3-branes can still recombine into D3-branes at the origin of the ALE space. For this reason they do not factorize into Higgs and Coulomb branch contributions.

\subsubsection{Cancellations with surgery}

We now want to express the terms appearing in (\ref{altsum}) in terms of simpler building blocks. Specifically, we want to prevent fractional D3-branes from recombining into D3-branes. 
It is not hard to convince oneself of the surgical decompositions
\beq
\raisebox{-.15in}{\includegraphics[width=1.02in]{graph21.eps}}=
\raisebox{-.15in}{\includegraphics[width=1.02in]{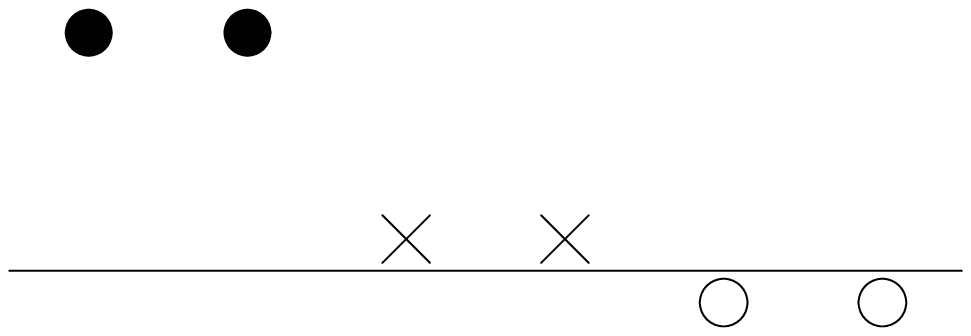}}-
\raisebox{-.15in}{\includegraphics[width=.85in]{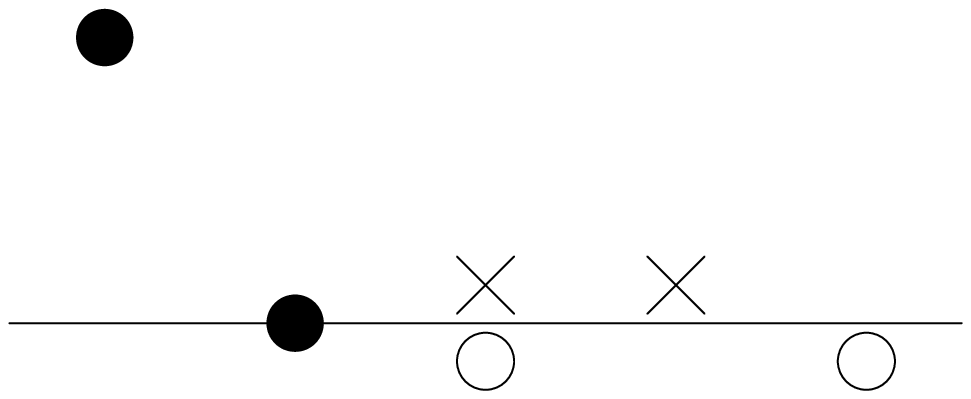}}+
\raisebox{-.15in}{\includegraphics[width=.85in]{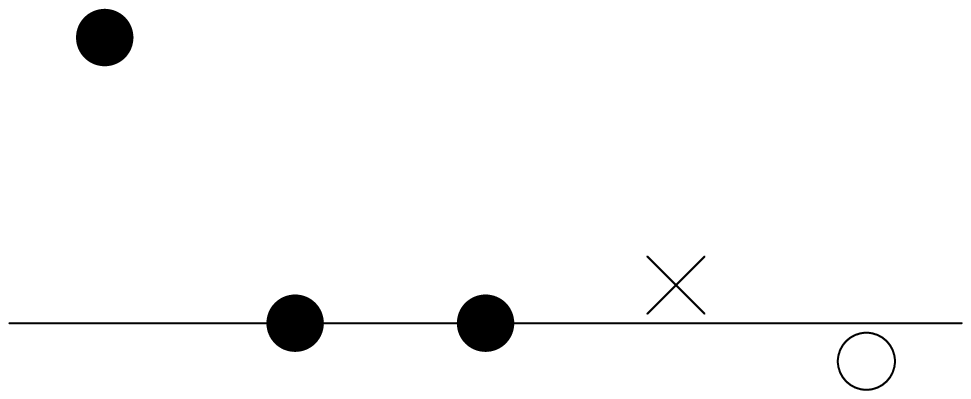}}-
\raisebox{-.15in}{\includegraphics[width=.68in]{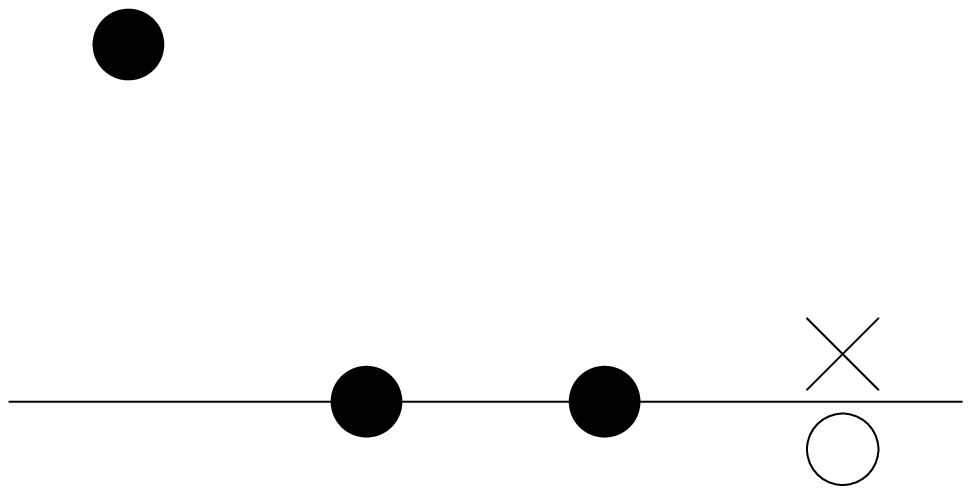}}+
\raisebox{-.15in}{\includegraphics[width=.68in]{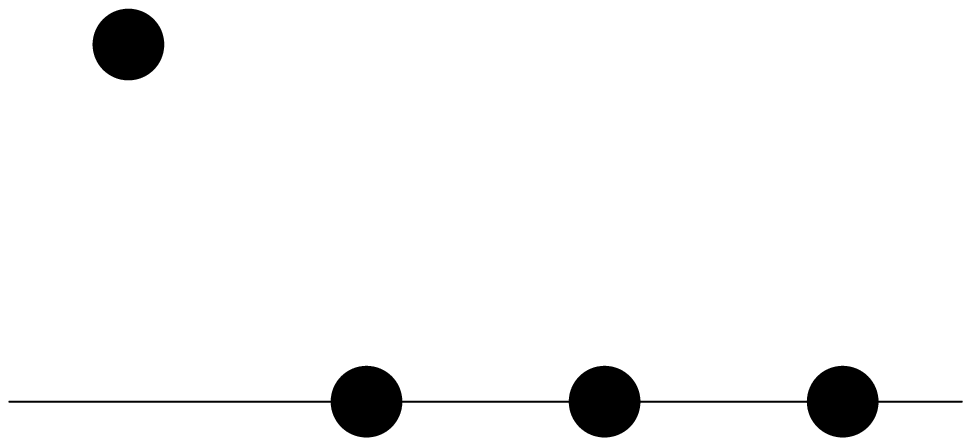}}
\eeq
and
\beq
\raisebox{-.15in}{\includegraphics[width=1.02in]{graph27.eps}}=
\raisebox{-.15in}{\includegraphics[width=1.02in]{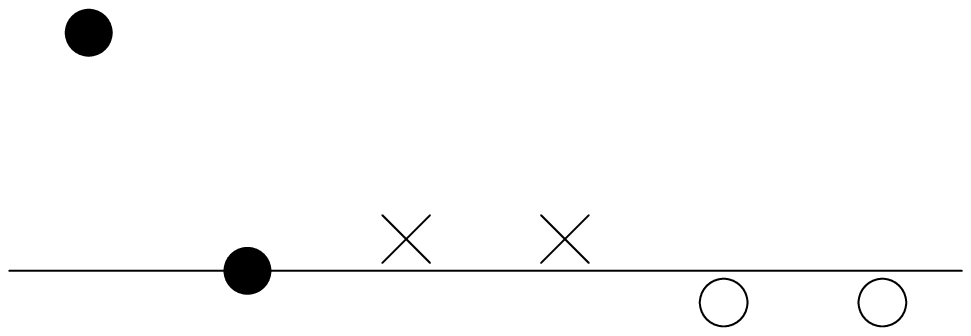}}-
\raisebox{-.15in}{\includegraphics[width=.85in]{graph23.eps}}+
\raisebox{-.15in}{\includegraphics[width=.85in]{graph24.eps}}-
\raisebox{-.15in}{\includegraphics[width=.68in]{graph25.eps}}+
\raisebox{-.15in}{\includegraphics[width=.68in]{graph26.eps}}
\eeq
%
etc.
Here the crosses and circles being on different sides of the horizontal axis indicates that the fractional D3-branes cannot recombine. A cross and a circle being at the same position along the horizontal axis indicate that the two fractional D3-branes are forced to sit at the same position.

In the alternating sum (\ref{altsum}) most of the terms cancel and only a few remain
\beq\label{naltsum}
g=\raisebox{-.15in}{\includegraphics[width=.51in]{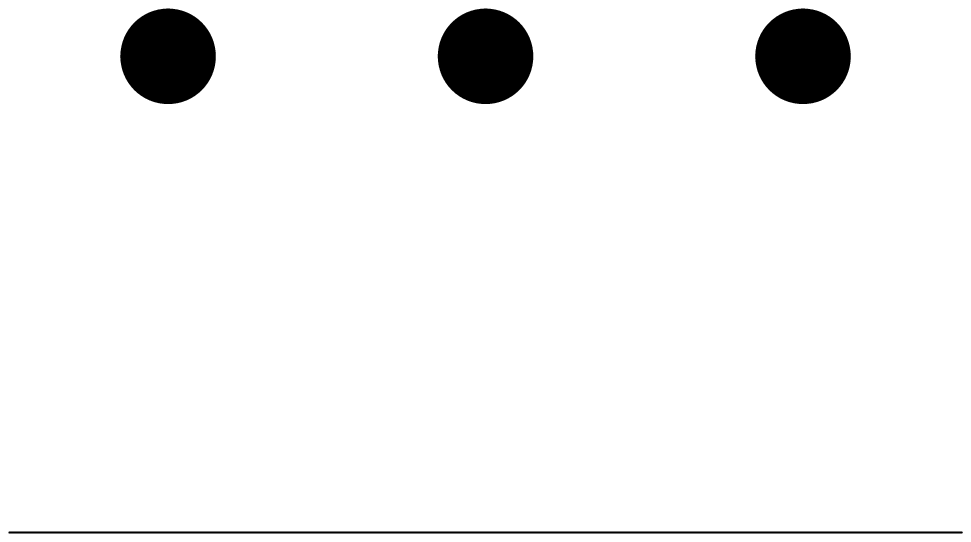}}
-\raisebox{-.15in}{\includegraphics[width=.51in]{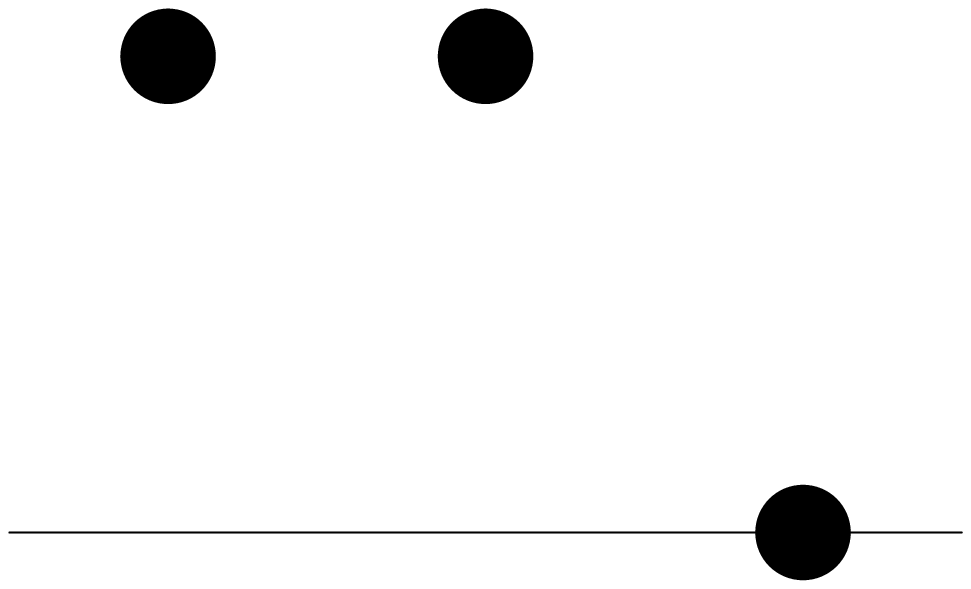}}
+\raisebox{-.15in}{\includegraphics[width=.68in]{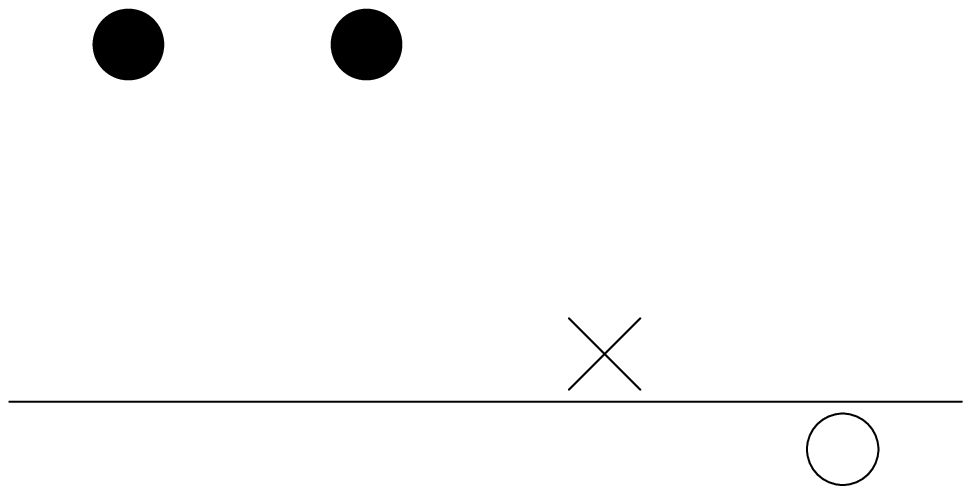}}
-\raisebox{-.15in}{\includegraphics[width=.68in]{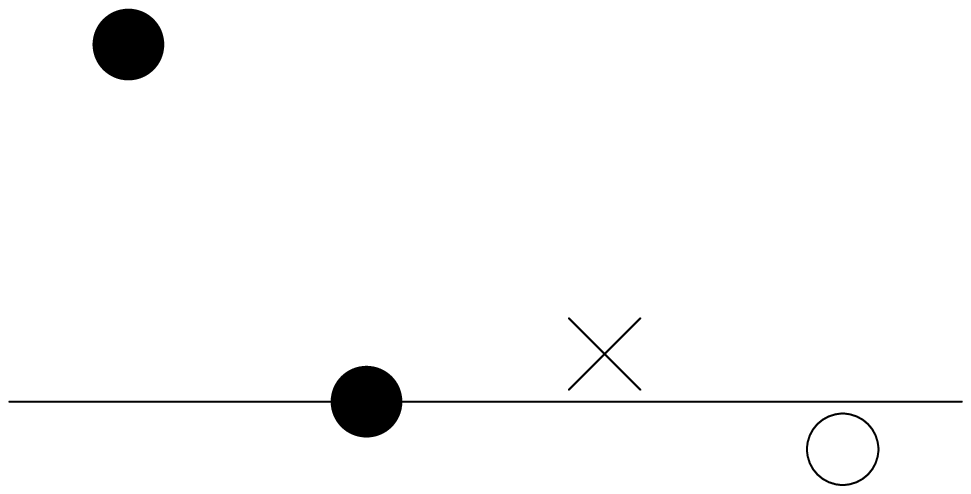}}
+\raisebox{-.15in}{\includegraphics[width=.85in]{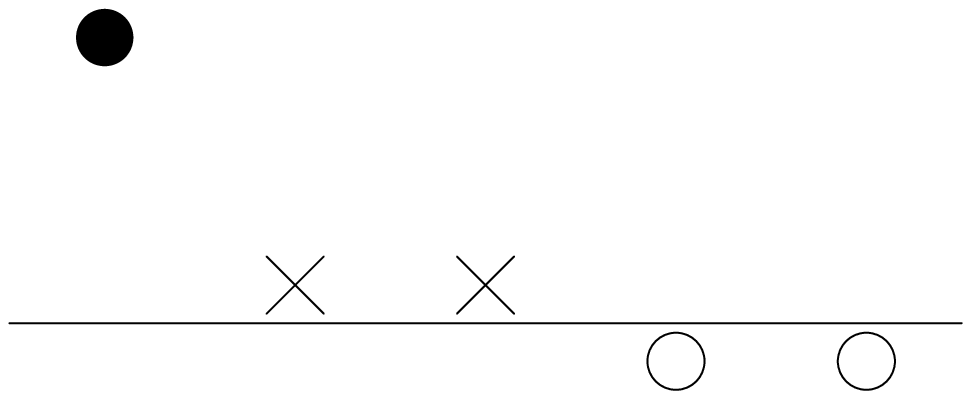}}
-\raisebox{-.15in}{\includegraphics[width=.85in]{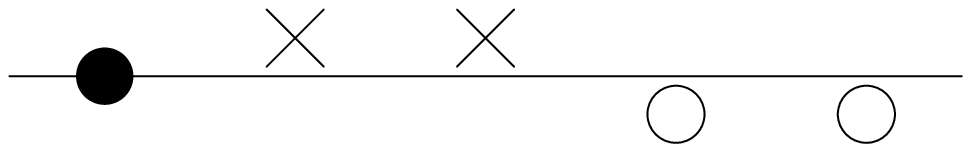}}
+\raisebox{-.15in}{\includegraphics[width=1.02in]{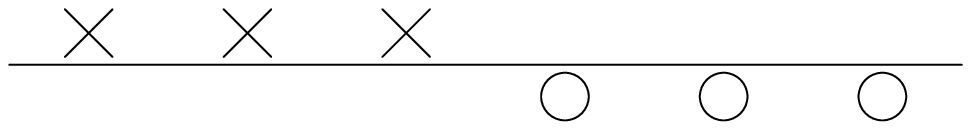}}
\eeq
The terms in this new alternating sum factorize into contributions of three types: Higgs, Coulomb and restricted Higgs.
\beq
\raisebox{-.15in}{\includegraphics[width=.6in]{graph31.eps}}\quad,\qquad
\raisebox{-.15in}{\includegraphics[width=1.2in]{graph37.eps}}\qquad{\rm and}\qquad
\raisebox{-.15in}{\includegraphics[width=.6in]{graph32.eps}}
\eeq
The first two types were discussed in sections \ref{higgs} and \ref{coulomb}.  Using generating functions equation (\ref{naltsum}) can be written as
\beq\label{gnaltsum}
G ( \nu ; t_1,t_2,t_3; \Gamma)=H ( \nu ; t_1,t_2,t_3; \Gamma)C ( \nu ; t_3; \Gamma)-(R ( \nu ; t_1,t_2,t_3; \Gamma)-1)C ( \nu ; t_3; \Gamma),
\eeq
where $R$ is the generating function for the restricted Higgs branch
\beq
R ( \nu ; t_1,t_2,t_3; \Gamma) = \sum_{N=0}^\infty \nu^N r_N (t_1,t_2,t_3; \Gamma).
\eeq

\subsubsection{More surgery}

To deal with a restricted Higgs branch we need to do more surgery. It is not hard to derive a recursion relation for the contributions from the restricted branch:
\beq
\raisebox{-.15in}{\includegraphics[width=.6in]{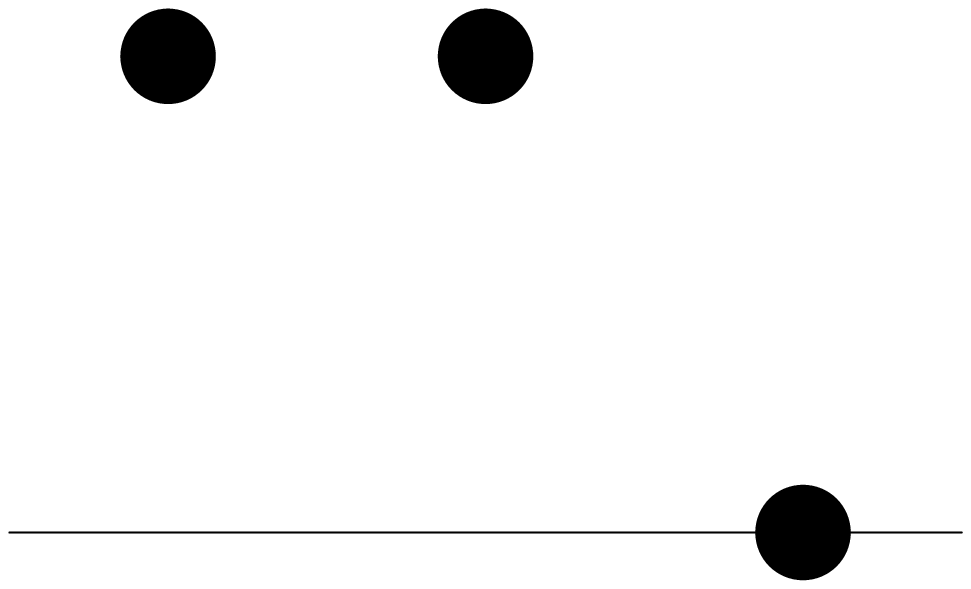}}=
\raisebox{-.15in}{\includegraphics[width=.6in]{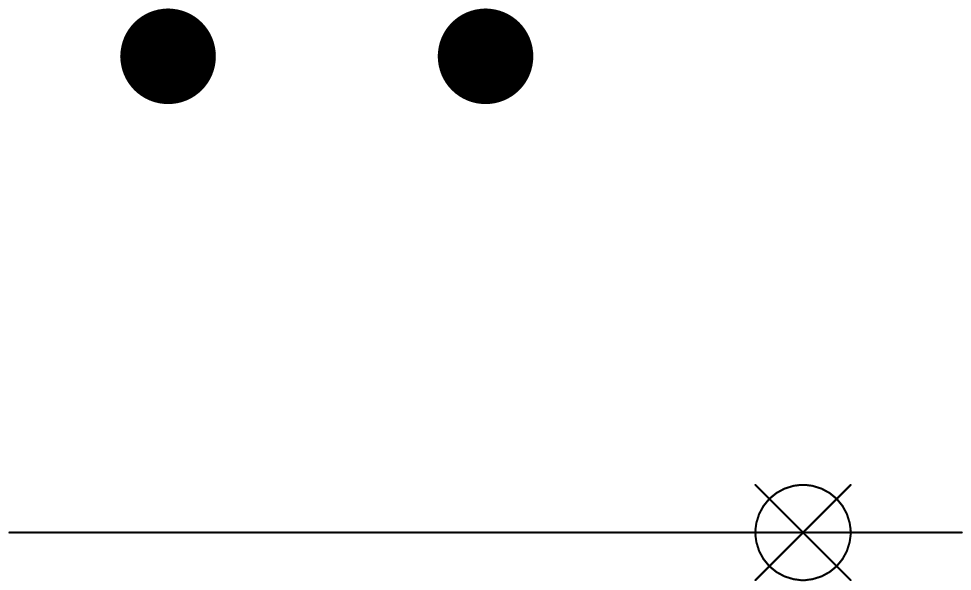}}-
\raisebox{-.15in}{\includegraphics[width=.6in]{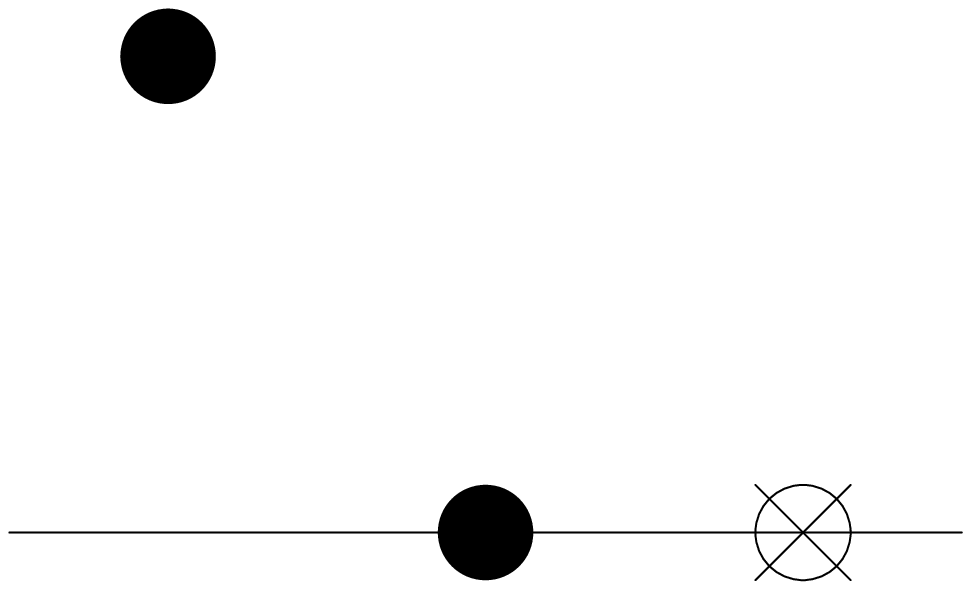}}+
\raisebox{-.15in}{\includegraphics[width=.6in]{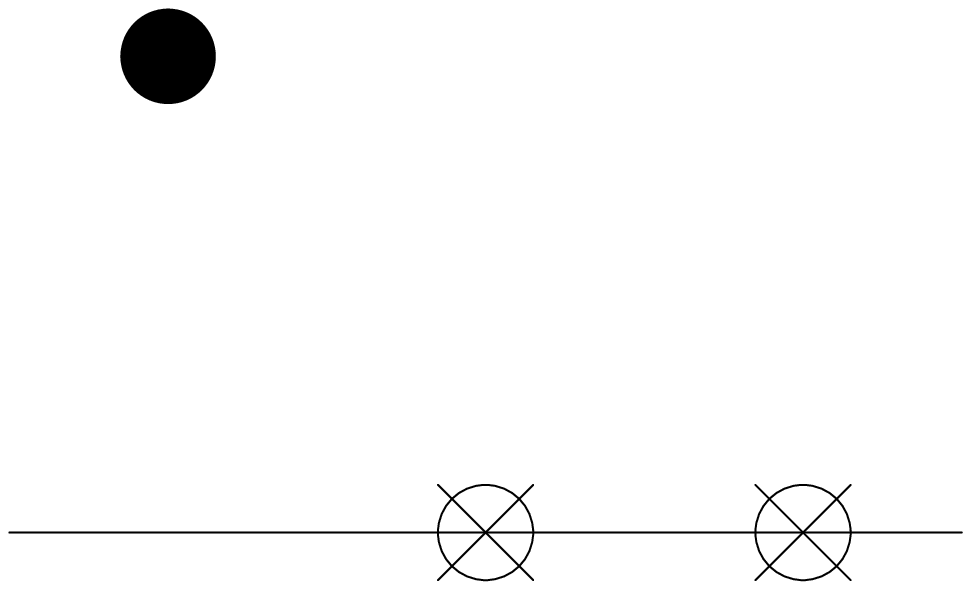}}-
\raisebox{-.15in}{\includegraphics[width=.6in]{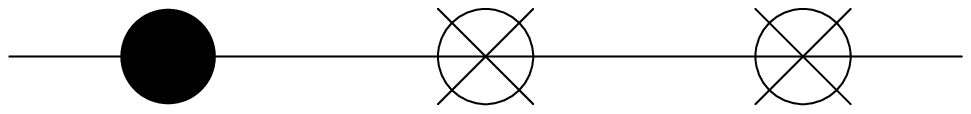}}+
\raisebox{-.15in}{\includegraphics[width=.6in]{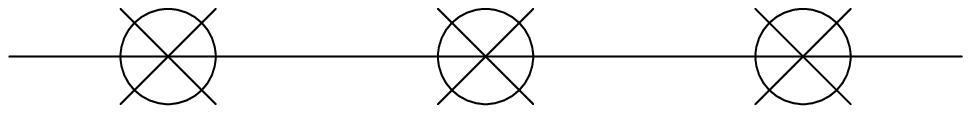}}
\eeq
In those diagrams a crossed circle on the horizontal axis represents a brane that is stuck at the origin of the ALE space, but can be distinguished from a D3-brane (filled dot) that is stuck at the origin of the ALE space. The generating function for those branes is $L ( \nu ; t_3) $.

Using the generating functions the above recursion relation can be written as
\beq\label{recrel}
H ( \nu ; t_1,t_2,t_3; \Gamma) (L ( \nu ; t_3) -1)=(R ( \nu ; t_1,t_2,t_3; \Gamma) -1)L ( \nu ; t_3).
\eeq
combining equations (\ref{gnaltsum}) and (\ref{recrel}) leads to the desired result
\beq\label{finresult}
G ( \nu ; t_1,t_2,t_3; \Gamma) =\frac{H ( \nu ; t_1,t_2,t_3; \Gamma) C ( \nu ; t_3; \Gamma) }{L ( \nu ; t_3) }.
\eeq

\section{Examples and a full solution to a class of gauge theories}
\label{examples}

\subsection{First few cases for small number of D3 branes}

We can use the expression for $G ( \nu ; t_i ; \Gamma )$, the full generating function of chiral operators 
to make general observations on the first few cases of small $N$.

We first observe that $h_0 = c_0 = l_0 = 1$, from equations (\ref{Hnu}), (\ref{Cnu}) and (\ref{Lnu}). This implies that
\beq
g_0 = 1.
\eeq
The moduli space of this theory is a point and there is only the identity operator in the chiral ring.

Second, using equation (\ref{Mnu}), we find
\beq
g_1 =  c_1 + h_1 - l_1.
\eeq
This equation expresses the simplest notions about ``surgery" as discussed in section \ref{surgery} using the simplest example of the complex equation $xy=0$. The mixed branch for one D3 brane gets contributions from:
\begin{enumerate}
\item  The Higgs branch, where the D3 brane can propagate as a physical D3 brane.
\item The Coulomb branch where the D3 brane splits into fractional branes which are propagating along the singularity.
\item The intersection between these two moduli spaces needs to be subtracted to avoid double counting of operators.
\end{enumerate}
Using the explicit expressions for the case of $\Z_n$ we find
\beq
g_1= \frac{1-t_1^nt_2^n}{(1-t_1^n)(1-t_2^n)(1-t_1t_2)(1-t_3)} + \frac{1}{(1-t_3)^n} - \frac{1}{(1-t_3)}
\eeq

Finally, using equation (\ref{Mnu}), we find
\beq\label{g2exp}
g_2 = c_2 + c_1h_1 + h_2 - (c_1 + h_1) l_1 + l_1^2 - l_2.
\eeq
For $A_1$ quiver gauge theories this result can also be checked in an elementary way by counting operators:

Starting with the alternating sum (\ref{altsum}) and noting that $b_2=h_2$ and $b_0 = c_2$ we have
\beq
g_2=b_2-i_2+b_1-i_1+b_0=h_2-i_2+b_1-i_1+c_2
\eeq
we therefore need to show that 
\beq
i_2-b_1+i_1=l_2+(c_1+h_1)l_1-c_1h_1-l_1^2.
\eeq
Let us start by constructing $b_1$, the contribution of the mixed branch.  The D and F flatness constraints imply that the $\phi_i$, $A_i$ and $B_i$, $i=1,2$ can be diagonalized and have the form
\beq
\phi_i=\left(\begin{array}{cc}\lambda_0&0\\0&\lambda_i\end{array}\right),\qquad
A_i=\left(\begin{array}{cc}a_i&0\\0&0\end{array}\right)\qquad{\rm and}\qquad
B_i=\left(\begin{array}{cc}b_i&0\\0&0\end{array}\right).
\eeq
The gauge invariant operators are then polynomials in $\lambda_0$, $\lambda_i$, $a_i$ and $b_i$.
\beqa
\label{trp}
\tr\phi_i&=&\lambda_0+\lambda_i,\\ \label{trps}
\tr\phi_i^2&=&\lambda_0^2+\lambda_i^2,\\ 
\tr\phi_i^mA_iB_i&=&\lambda_0^ma_ib_i\qquad{\rm etc.}
\eeqa
Equations \eqref{trp} and \eqref{trps} have three independent parameters. We could try to generate all operators containing only $\phi_i$ by $\tr\phi_i$ and $\tr\phi_i^2$. These are four variables and therefore we need to introduce one constraint which turns out to be quartic. This implies that they are counted by the generating function
\beq
\frac{1-t_3^4}{(1-t_3)^2(1-t_3^2)^2}.
\eeq
Operators that contain only $A_i$ and $B_i$ are counted by
\beq
\frac{1+t_1t_2}{(1-t_1^2)(1-t_2^2)}.
\eeq
Finally, it is not hard to see that operators that contain at least one $A_i$ or $B_i$ can have any polynomial in $\lambda_0$, $\lambda_1$ and $\lambda_2$ in front of it, i.e. they are counted by
\beq
\left(\frac{1+t_1t_2}{(1-t_1^2)(1-t_2^2)}-1\right)\frac{1}{(1-t_3)^3}.
\eeq
This implies that
\beq
b_1=\left(\frac{1+t_1t_2}{(1-t_1^2)(1-t_2^2)}-1\right)\frac{1}{(1-t_3)^3}+\frac{1-t_3^4}{(1-t_3)^2(1-t_3^2)^2}.
\eeq

Similarly, one can derive
\beq
i_1=\frac{1-t_3^4}{(1-t_3)^2(1-t_3^2)^2}
\eeq
and
\beq
i_2=\left(\frac{1+t_1t_2}{(1-t_1^2)(1-t_2^2)}-1\right)\frac{1}{(1-t_3)^2}+\frac{1}{(1-t_3)(1-t_3^2)}.
\eeq

This implies that
\beq
i_2-b_1+i_1=\left(1-\frac{1+t_1t_2}{(1-t_1^2)(1-t_2^2)}\right)\frac{t_3}{(1-t_3)^3}+\frac{1}{(1-t_3)(1-t_3^2)}=l_2+(c_1+h_1)l_1-c_1h_1-l_1^2,
\eeq
which proves (\ref{g2exp}) in a more elementary way.

\subsection{The $A_n$ series}

The generating function of the Higgs branch was calculated in \cite{Benvenuti:2006qr} and takes the form
\beq
g_1(t_1,t_2,t_3; \BC^2/\Z_n\times \BC) = \frac{1-t_1^nt_2^n}{(1-t_1^n)(1-t_2^n)(1-t_1t_2)(1-t_3)},
\eeq
Taking the plethystic exponential of this function gives the generating function for multi trace operators and rank of the gauge group equal to $N$,
\beq
H(\nu; t_1, t_2, t_3; \BC^2/\Z_n\times \BC) = \exp\biggl(\sum_{k=1}^\infty\frac{\nu^kg_1(t_1^k,t_2^k,t_3^k)}{k}\biggr)
\eeq
The generating function on the Coulomb branch is given by
\beq
C(\nu;  t; \BC^2/\Z_n\times \BC) = \sum_{k=0}^\infty\nu^k\prod_{i=1}^k\frac{1}{(1-t^i)^n}.
\eeq
The generating function of the line is given by
\beq
L(\nu; t) = \exp\biggl(\sum_{k=1}^\infty\frac{\nu^k}{k(1-t^k)}\biggr).
\eeq
Finally, the full generating function is given by
\beqa
& &G(\nu; t_1, t_2, t_3; \BC^2/\Z_n\times \BC) = \frac { H(\nu; t_1, t_2, t_3; \BC^2/\Z_n\times \BC) C(\nu;  t_3; \BC^2/\Z_n\times \BC) } { L(\nu; t_3) } = \nonumber \\ &=& \exp\biggl[\sum_{k=1}^\infty\frac{\nu^k}{k(1-t_3^k)}\biggl(
\frac{1-t_1^{nk}t_2^{nk}}{(1-t_1^{nk})(1-t_2^{nk})(1-t_1^kt_2^k)}-1\biggr)\biggr] \sum_{j=0}^\infty\nu^j\prod_{i=1}^j\frac{1}{(1-t^i)^n} \nonumber
\eeqa





\section{Conclusions}

In this paper we studied the generating function that counts BPS operators in the chiral ring of a ${\cal N}=2$ quiver gauge theory which arises on the world volume of D3 branes probing an ALE singularity in Type IIB superstring theory. The difficulty in finding the generating function lies in the presence of mixed Coulomb and Higgs Branches and careful analysis of the counting needs to be done. Luckily we develop ``surgery" techniques which allow the precise calculation of the generating functions and in fact give a very simple final result, equation \eqref{Mnu}.

The problem of counting BPS operators in the chiral ring of a generic supersymmetric gauge theory remains elusive. Some recent progress has been made in incorporating baryonic charges in dimer theories \cite{Butti:2006au} and some progress in understanding theories with no superpotential. More results are hoped to be reported in the near future.

The behavior of the generating functions at large values of the charges is of importance to evaluate the entropy of these gauge theories and will be deferred to a future publication.

\begin{acknowledgments} 

\noindent Research at the Perimeter Institute is supported in part by funds from NSERC of Canada.

A.~H.~ would like to thank Sergio Benvenuti, Bo Feng, Guido Festuccia, Jaume Gomis, and Yang-Hui He for very fruitful discussions. C.~R.~would like to thank Sergey Cherkis for very fruitful discussions.

\end{acknowledgments}

\vfill
\pagebreak

\bibliographystyle{JHEP}

 \providecommand{\href}[2]{#2}\begingroup\raggedright\endgroup

\end{document}